\documentclass[12pt,thmsa]{article}
\usepackage{amsfonts}

\usepackage{sw20lart}

\input{tcilatex}

\begin{document}

\author{Claudio Garola}
\title{Embedding QM into an objective framework}
\date{Dipartimento di Fisica dell'Universit\`{a} and Sezione INFN, 73100 Lecce,
Italy. E-mail: Garola@le.infn.it}
\maketitle

\begin{abstract}
An elementary model is given which shows how an objective (hence local and
noncontextual) picture of the microworld can be constructed without
conflicting with quantum mechanics (QM). This contradicts known no-go
theorems, which however do not hold in the model, and supplies some
suggestions for a broader theory in which QM can be embedded.

\smallskip

\end{abstract}

According to the standard (Copenhagen) interpretation, quantum mechanics
(QM) is \textit{nonobjective}, which can be briefly expressed by saying that
``a measurement does not, in general, reveal a preexisting value of the
measured property''$^{(1)}$. Though accepted by generations of physicists,
nonobjectivity implies a number of nonintuitive consequences and puzzling
paradoxes, which spread out from QM to all theories based on it. However, it
is strongly supported by a number of arguments, among which
Bell-Kochen-Specker's$^{(2),(3)}$ and Bell's$^{(4),(5)}$. These are usually
seen as no-go theorems which show that a noncontextual and local (hence
objective) picture of the microworld cannot be consistent with QM, so that
one must come to terms with the paradoxes following from nonobjectivity.

The elementary set-theoretical model provided here aims to show that the
above conclusion can be circumvented without altering the formalism and the
minimal (statistical) interpretation of QM. The resulting objective picture,
despite its simplicity, has some deep theoretical implications, that will be
briefly expounded and commented on at the end.

To begin with, let us accept the standard notion of state of a physical
system $\frak{S}$ as a class of physically equivalent preparing devices.$%
^{(6)}$ Furthermore, let us call \textit{physical object} any individual
sample x of $\frak{S}$ obtained by activating a preparing device, and say
that \textit{x is in the state S} (at time t) if the device $\pi $ preparing
x (at t) belongs to S. Whenever $\frak{S}$ is a microscopic physical system,
let us introduce a set $\mathcal{E}$ of \textit{microscopic} physical
properties \textit{f}, \textit{g}, ..., characterizing $\frak{S}$, which
play the role of theoretical entities. For every physical object x, every
property \textit{f }$\in \mathcal{E}$ is associated with x in a dichotomic
way, so that one briefly says that every \textit{f }$\in \mathcal{E}$ either
is possessed or it is not possessed by x. The set $\mathcal{F}_{0}$ of all 
\textit{macroscopic} properties is then introduced as in standard QM, that
is, it is defined as the set of all pairs of the form ($\mathcal{A}_{0}$,$%
\Delta $), where $\mathcal{A}_{0}$ is an observable (that is, a class of
physically equivalent measuring apparatuses) with spectrum $\Lambda _{0}$,
and $\Delta $ a Borel set on the real line (for every observable $\mathcal{A}%
_{0}$, different sets containing the same subset of $\Lambda _{0}$ obviously
define physically equivalent properties). Yet, every observable $\mathcal{A}%
_{0}$ is obtained from a suitable observable $\mathcal{A}$ of standard QM by
adding to the spectrum $\Lambda $ of $\mathcal{A}$ a further outcome a$_{0}$
that does not belong to $\Lambda $, called the \textit{no-registration}
outcome of $\mathcal{A}_{0}$ (note that such an outcome can be introduced
also within the standard quantum measurement theory, but it plays here a
different theoretical role), so that $\Lambda _{0}$ = $\Lambda \cup \left\{ 
\text{a}_{0}\right\} $. The set $\mathcal{E}$ of all microscopic properties
is then assumed to be in one-to one correspondence with the subset $\mathcal{%
F}$ $\mathcal{\subseteq }$ $\mathcal{F}_{0}$\ of all macroscopic properties
of the form F = ($\mathcal{A}_{0}$,$\Delta $), where $\mathcal{A}_{0}$ is an
observable and a$_{0}\notin \Delta $.

Basing on the above definitions and assumptions, one can provide the
following description of the measurement process. Whenever a physical object
x is prepared by a given device $\pi $ in a state S, and $\mathcal{A}_{0}$
is measured by means of a suitable apparatus, the set of microscopic
properties possessed by x produces a probability (which is either 0 or 1 if
the model is \textit{deterministic}) that the apparatus does not react, so
that the outcome a$_{0}$ may be obtained. In this case, x is not detected
and one cannot get any explicit information about the microscopic physical
properties possessed by x. If, on the contrary, the apparatus reacts, an
outcome different from a$_{0}$, say a, is obtained, and one is informed that
x possesses all microscopic properties associated with macroscopic
properties of the form F = ($\mathcal{A}_{0}$,$\Delta $), where $\Delta $ is
a Borel set such that a$_{0}\notin \Delta $ and a $\in \Delta $ (for the
sake of brevity we also say that x \textit{possesses} all macroscopic
properties as F in this case).

In order to place properly quantum probability within the above intuitive
picture, let us consider a preparing device $\pi \in $ S that is activated
repeatedly. In this case a (finite) set $\mathcal{S}$ of physical objects in
the state S is prepared. Then, let us partition $\mathcal{S}$ into subsets $%
\mathcal{S}^{(1)}$, $\mathcal{S}^{(2)}$, ..., $\mathcal{S}^{(n)}$, such that
in each subset all objects possess the same \textit{microscopic} properties,
and assume that a measurement of an observable $\mathcal{A}_{0}$ is done on
every object. Furthermore, let us introduce the following symbols.

N: number of physical objects in $\mathcal{S}$.

N$_{0}$: number of physical objects in $\mathcal{S}$ that are not detected.

N$^{(i)}$: number of physical objects in $\mathcal{S}^{(i)}$.

N$_{0}^{(i)}$: number of physical objects in $\mathcal{S}^{(i)}$ that are
not detected.

N$_{F}^{(i)}$: number of physical objects in $\mathcal{S}^{(i)}$ that
possess the macroscopic property F = ($\mathcal{A}_{0}$,$\Delta $)
corresponding to the microscopic property \textit{f}.

It is apparent that the number N$_{F}^{(i)}$ either coincides with N$^{(i)}-$%
N$_{0}^{(i)}$ or with 0. The former case occurs whenever \textit{f} is
possessed by the objects in $\mathcal{S}^{(i)}$, since all objects that are
detected then yield outcome in $\Delta $. The latter case occurs whenever 
\textit{f} is not possessed by the objects in $\mathcal{S}^{(i)}$, since all
objects that are detected then yield outcome different from a$_{0}$ but
outside $\Delta $. In both cases one generally gets N$^{(i)}-$N$_{0}^{(i)}$ $%
\neq $ 0 (even if N$^{(i)}-$N$_{0}^{(i)}$ = 0 may also occur, in particular
in a deterministic model), so that the following equation holds:

\begin{description}
\item  $\frac{N_{F}^{(i)}}{N^{(i)}}=\frac{N^{(i)}-N_{0}^{(i)}}{N^{(i)}}\frac{%
N_{F}^{(i)}}{N^{(i)}-N_{0}^{(i)}}$ .\hspace{3.05in}(1)
\end{description}

The term on the left in Eq. (1) represents the frequency of objects
possessing the property F in $\mathcal{S}^{(i)}$, the first term on the
right the frequency of objects in $\mathcal{S}^{(i)}$ that are detected, the
second term (which either is 1 or 0) the frequency of objects that possess
the property F in the subset of all objects in $\mathcal{S}^{(i)}$ that are
detected.

The frequency of objects in $\mathcal{S}$ that possess the property F is
given by

\begin{description}
\item  $\frac{1}{N}\sum_{i}N_{F}^{(i)}=\frac{N-N_{0}}{N}(\sum_{i}\frac{%
N_{F}^{(i)}}{N-N_{0}})$ .\hspace{2.7in}(2)
\end{description}

Let us assume now that all frequencies converge in the large number limit,
so that they can be substituted by probabilities, and that these
probabilities do not depend on the choice of the preparing device $\pi $\ in
S. Hence, if one considers the large number limit of Eq. (1), one gets

\begin{description}
\item  $\mathcal{P}_{S}^{(i)t}(F)=\mathcal{P}_{\mathcal{S}}^{(i)d}(F)%
\mathcal{P}_{S}^{(i)}(F)$ ,\hspace{2.87in}(3)
\end{description}

\noindent where $\mathcal{P}_{S}^{(i)t}(F)$ is the overall probability that
a physical object x possessing the microscopic properties that characterize $%
\mathcal{S}^{(i)}$ also possess the property F, $\mathcal{P}_{S}^{(i)d}(F)$
is the probability that x be detected when F is measured on it, $\mathcal{P}%
_{S}^{(i)}(F)$ (which either is 0 or 1) is the probability that x possess
the property F when detected. Moreover, the large number limit of Eq. (2)
yields

\begin{description}
\item  $\mathcal{P}_{S}^{t}(F)=\mathcal{P}_{\mathcal{S}}^{d}(F)\mathcal{P}%
_{S}(F)$ ,\hspace{3.1in}(4)
\end{description}

\noindent where $\mathcal{P}_{S}^{t}(F)$ is the overall probability that a
physical object x in a state S possess the property F, $\mathcal{P}%
_{S}^{d}(F)$ is the probability that x be detected when F is measured on it, 
$\mathcal{P}_{S}(F)$ is the probability that x possess the property F when
detected. It is then reasonable to identify $\mathcal{P}_{S}(F)$ with the
quantum probability that a physical object in the state S possess the
property F, so that $\mathcal{P}_{S}(F)$ can be evaluated by following the
rules of standard QM, hence in particular representing any state S by means
of a trace class operator on a Hilbert space $\mathcal{H}$ associated with $%
\frak{S}$ and any macroscopic property that corresponds to a microscopic
property by means of a projection operator on $\mathcal{H}$. Thus, one need
not modifying the formalism and the statistical interpretation of standard
QM.

As anticipated at the beginning, however, the set-theoretical model
illustrated above provides an objective (hence local and noncontextual)
picture of the microworld which is consistent with QM. Indeed, for every
physical object x in the state S, every macroscopic property of the form F =
($\mathcal{A}_{0}$,$\Delta $) (where a$_{0}$ may belong or not to $\Delta $)
either is possessed or is not possessed by x, and the probability that it is
possessed/not possessed is determined by the microscopic properties
possessed by x, which do not depend on the measuring apparatus (hence
microscopic properties play in the model a role similar to states in
objective local theories$^{(7)}$). This violates standard expectations and
can be explained as follows.

The Hilbert space formalism of standard QM does not associate any
mathematical object with the microscopic properties \textit{f}, \textit{g},
... Furthermore, projection operators represent only macroscopic properties
of the form ($\mathcal{A}_{0}$,$\Delta $), where $\mathcal{A}_{0}$ is an
observable and a$_{0}\notin \Delta $, so that the mathematical
representation of the entities appearing in the model is only partial.
Hence, every QM law stated by means of the standard formalism necessarily
relates (possibly probabilistically) only entities that are mathematically
represented, that is, states and macroscopic properties of the form
specified above. If the law is interpreted in the observative language of
the theory, it may undergo a process of empirical verification, and one can
classify it as \textit{empirical}. Yet, because of the above remarks, such a
law refers only to objects that are detected; moreover, it can be actually
verified only in those physical situations in which the verification
procedure does not lead to simultaneous measurements of noncommeasurable
observables. If these restrictive conditions are satisfied, the relations
among macroscopic properties established by the law match analogous
relations among the corresponding microscopic properties. If, on the
contrary, the restrictive conditions are not satisfied, one can neither
assert the validity of the relations predicted by the law among macroscopic
properties, nor transfer these relations to microscopic properties. As an
example, think of a physical object x on which an observable $\mathcal{A}%
_{0} $ is measured, obtaining outcome a$_{0}$. In this case, no macroscopic
property of the form ($\mathcal{A}_{0}$,$\Delta $), with a$_{0}\notin \Delta 
$, is possessed by x, hence no non-trivial relation among properties of this
form holds, and no relation among the microscopic properties possessed by x
can be inferred from quantum laws (but the model predicts that microscopic
properties must be such that the probability of the a$_{0}$ outcome is not
0).

The above arguments point out that quantum laws must be handled with care
within the model. In particular, consider the condition stated by Kochen and
Specker (briefly, KS)$^{(3)}$ as a basic premise for the Bell-KS theorem,
that can be reformulated as follows.$^{(1)}$

\textit{If a set of mutually commuting observables A, B, C, ... satisfies a
relation of the form f(A, B, C, ...) = 0 then the values v(A), v(B), v(C),
... assigned to them in an individual system must also be related by}

\begin{description}
\item  \textit{f(v(A), v(B), v(C), ...) = 0 }.\hspace{3.1in}(5)
\end{description}

In all proofs of the Bell-KS theorem, the law (5) is applied repeatedly,
inserting in it different sets of mutually commuting observables, and there
are observables belonging to different sets that do not commute. This
implies that, if Eq. (5) is checked for a given choice of observables,
checking it (on the same objects) for a different choice requires
simultaneous measurements of noncommeasurable observables. Hence, one cannot
assert that all relations among macroscopic properties established by
equations of the form (5) are bound to hold simultaneously, nor that they
can be translated into relations among the corresponding microscopic
properties. This invalidates the premises on which the Bell-KS theorem
stands, which explains how the model can circumvent this theorem and provide
a noncontextual picture of the microscopic world.

Similar reasonings apply if Bell's inequalities are considered. These are
usually maintained to show that QM is a nonlocal theory (Bell's theorem).
Let us refer, for the sake of simplicity, to the inequalities propounded by
Clauser \textit{et al}. in 1969 (\textit{CHSH's inequalities}).$^{(5)}$
Then, the quantum inequalities corresponding to CHSH's inequalities relate
(dichotomic) observables, hence macroscopic properties, and can be checked.
However, the check is not trivial, since the inequalities contain
noncommeasurable observables, so that they can be checked only ``by
blocks'', that is, measuring different correlation functions on different
sets of physical objects, all in the same state. But this procedure
considers in every set only the objects that are actually detected, and the
frequencies that are obtained must be interpreted in terms of probabilities
of the form $\mathcal{P}_{S}(F)$ (see Eq. 4), that are related as in QM, not
in terms of probabilities of microscopic properties that do not appear in
the formalism of QM. On the other side CHSH's inequalities can be
interpreted in the model as relating correlation functions of microscopic
properties possessed by the physical objects, hence need not coincide with
the corresponding quantum inequalities. This illustrates how the model can
circumvent the Bell theorem and provide a local picture of the microscopic
world.

The following remarks point out some further features of the model.

(i) From the viewpoint of the model, QM is a theory that is incomplete in
several senses (it does not provide the probabilities $\mathcal{P}%
_{S}^{t}(F) $ and $\mathcal{P}_{S}^{d}(F)$ in Eq. (4), it does not say
anything about the distribution of microscopic properties on physical
objects in a given state whenever the objects are not detected, etc.). This
agrees with the conclusion of Einstein, Podolski and Rosen in their famous
paper$^{(8)}$, which was however discarded by most physicists in favor of
the opposite thesis upholded by Bohr. The model thus shows that the EPR
perspective was not necessarily inconsistent with QM. If this is accepted, a
broader theory embodying QM can be envisaged, according to which the quantum
probability $\mathcal{P}_{S}(F)$ is considered as a conditional rather than
an absolute probability (see Eq. (4)). It is then interesting to note that
Eq. (4) could also be obtained in the framework of a model which introduces $%
\mathcal{P}_{\mathcal{S}}^{d}(F)$ as efficiency of a non-ideal measuring
apparatus, as in some existing attempts of rescuing local realism by
resorting to the low efficiencies of the apparatuses in the existing
experiments that confirm quantum inequalities (see, e.g., Refs. 9 and 10).
Yet, in a model of this kind the no-registration outcome occurs because of
flaws of the measuring apparatus, hence $\mathcal{P}_{\mathcal{S}}^{d}(F)$
is 1 if ideal observables are considered. In the model presented here,
instead, the no-registration outcome may occur because of the microscopic
properties of the physical object. Hence, $\mathcal{P}_{\mathcal{S}}^{d}(F)$
may be less than 1 also in the case of an ideal apparatus (indeed every $\pi 
$ in S prepares objects which do not possess the same microscopic
properties, and some objects may possess sets of properties that make the
detection of them by any apparatus measuring F possible but not certain, or
even impossible).

(ii) The microscopic properties \textit{f}, \textit{g}, ... are hidden
parameters in the model, but are not hidden variables in the standard sense.
Indeed, it has been shown above that they are not bound to make Eq. (5)
valid in every physical situation, which instead is required as a basic
condition in the standard definitions of hidden variables.$^{(1),(3)}$ This
explains why microscopic properties are not contextual, as standard hidden
variables must be.

Finally, note that the model presented here can be placed within the broader
context of the objective interpretation of QM propounded by the author (see,
e.g., Refs. 11-15). However, it is sufficient by itself to open some new
interesting possibilities, that are usually ignored (or maintained
impossible) in the literature.

\end{document}